\documentclass[aps,reprint,onecolumn,superscriptaddress]{revtex4-2} 
\usepackage{amsfonts}
\usepackage{amssymb}
\usepackage{amsthm,amsmath}
\usepackage{mathrsfs}
\usepackage{indentfirst}
\usepackage{multirow}
\usepackage{graphicx}
\usepackage{amsmath}
\usepackage[amssymb]{SIunits}
\usepackage{amssymb}
\usepackage{color}
\usepackage[usenames,dvipsnames]{xcolor}
\usepackage{natbib}
\usepackage[T1]{fontenc}
\usepackage[hidelinks, bookmarks=true]{hyperref}
\usepackage{bm}
\usepackage{booktabs}
\usepackage{multirow}
\usepackage{setspace}

\begin{document}

\title{How Brine Pockets Evolve and Release in Ice}
\author{Xinyu Tang}
\affiliation{New Cornerstone Science Laboratory, Center for Combustion Energy, Key Laboratory for Thermal Science and Power Engineering of MoE, Department of Energy and Power Engineering, Tsinghua University, China}

\author{Feng Wang}
\thanks{Corresponding author(s): fengwang2023@tsinghua.edu.cn; chaosun@tsinghua.edu.cn}
\affiliation{New Cornerstone Science Laboratory, Center for Combustion Energy, Key Laboratory for Thermal Science and Power Engineering of MoE, Department of Energy and Power Engineering, Tsinghua University, China}

\author{Yihong Du}
\affiliation{New Cornerstone Science Laboratory, Center for Combustion Energy, Key Laboratory for Thermal Science and Power Engineering of MoE, Department of Energy and Power Engineering, Tsinghua University, China}

\author{Yuki Wakata}
\affiliation{New Cornerstone Science Laboratory, Center for Combustion Energy, Key Laboratory for Thermal Science and Power Engineering of MoE, Department of Energy and Power Engineering, Tsinghua University, China}

\author{Jin Hu}
\affiliation{New Cornerstone Science Laboratory, Center for Combustion Energy, Key Laboratory for Thermal Science and Power Engineering of MoE, Department of Energy and Power Engineering, Tsinghua University, China}

\author{Chao Sun}
\thanks{Corresponding author(s): fengwang2023@tsinghua.edu.cn; chaosun@tsinghua.edu.cn}
\affiliation{New Cornerstone Science Laboratory, Center for Combustion Energy, Key Laboratory for Thermal Science and Power Engineering of MoE, Department of Energy and Power Engineering, Tsinghua University, China}
\affiliation{Department of Engineering Mechanics, School of Aerospace Engineering, Tsinghua University, China}

\date{\today}

\begin{abstract} 
Liquid inclusions migrate through solids by melting at one end and freezing at the other, but how a macroscopic phase boundary regulates their deformation and release remains poorly understood. Here, we track individual brine pockets in saline ice through bulk migration, elongation, and release at the ice–water interface, resolving the motion of both ends. In the bulk, temperature-dependent thermomigration gradually elongates the pockets. Near the ice--water interface, the front slows, whereas liquid connection is followed by accelerated rear freezing and rapid shortening. These opposing responses arise from distinct changes in thermal-solutal transport: proximity to the boundary suppresses the local thermal-gradient enhancement and slows front-end melting, while liquid connection enables salt exchange with the surrounding liquid and accelerates rear-end freezing. A quasi-steady diffusion-resistance model captures the dependence of the half-length release time on the pocket length at connection and identifies an effective external exchange length consistent with a salt-balance estimate. Our results identify thermal proximity and liquid connectivity as distinct controls on inclusion evolution and release, providing a framework for understanding liquid redistribution in multicomponent solids.
\end{abstract}

\maketitle
\section{Introduction}
\begin{figure*}[t]
\centering
\includegraphics[width=1\linewidth]{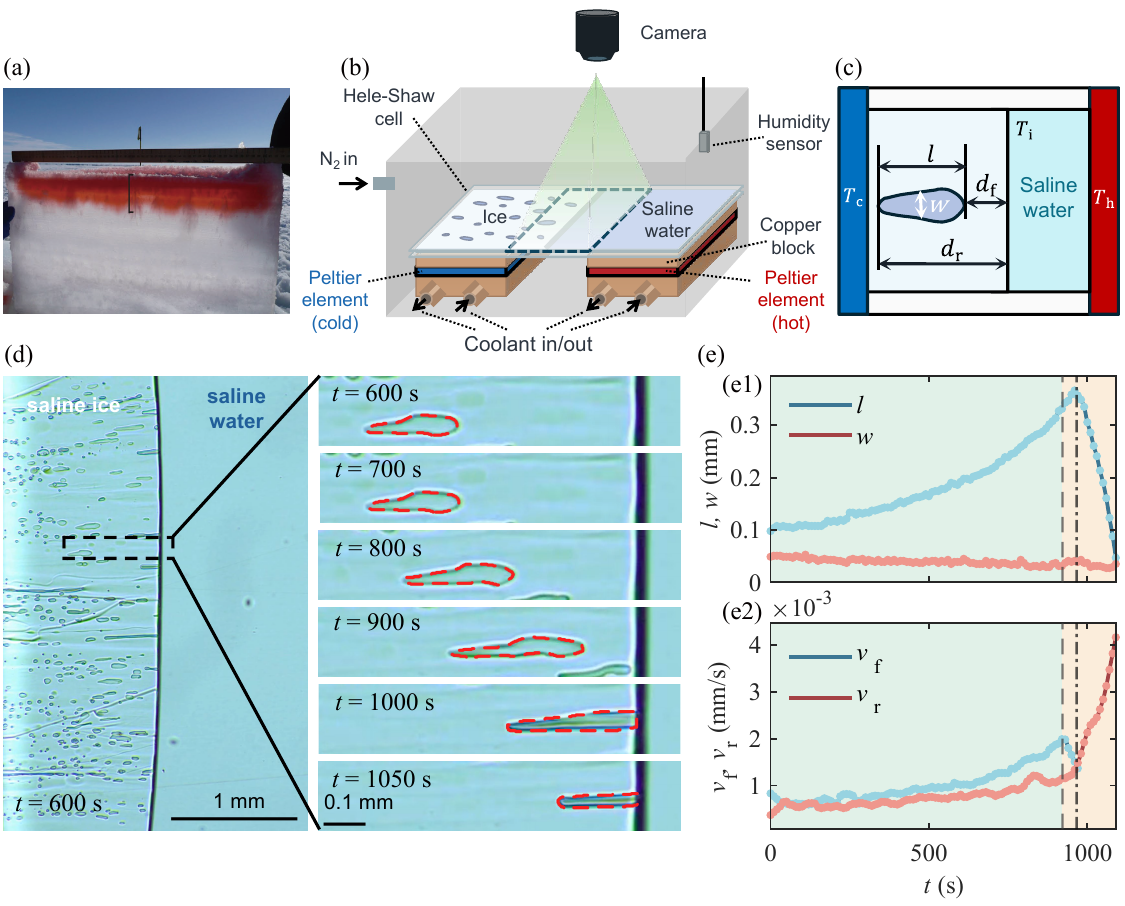}
\caption{Migration, elongation, and release of a brine pocket in saline ice. (a) Photograph of natural sea ice illustrating depth-dependent heterogeneity in brine volume fraction and pore connectivity, visualized by dye-tracer penetration. Reproduced from Golden et al. \cite{golden2026percolation} under CC BY-NC-ND 4.0. (b) Schematic of the temperature-controlled Hele–Shaw experimental setup. (c) Experimental geometry and definitions. A brine pocket of length $l$ and width $w$ migrates through an ice layer of mean thickness $L$ in a Hele-Shaw cell. $T_\mathrm{c}$, $T_\mathrm{h}$, and $T_\mathrm{i}$ denote the effective cold-side, hot-side, and ice--water interfacial temperatures, respectively. The distances from the front and rear tips to the local ice--water interface are denoted by $d_\mathrm{f}$ and $d_\mathrm{r}$, respectively, with $l=d_\mathrm{r}-d_\mathrm{f}$. (d) Experimental images of brine pocket migration. Left: the full-field image of brine pocket migration at $t=600\,\mathrm{s}$. The scale bar is 1 mm. The dashed rectangle indicates the zoomed-in region shown in the image on the right. Right: time-lapse images of the pocket in the enlarged region indicated on the left, showing the migration and evolution of a representative brine pocket at $T_\mathrm{c}=-3.3~^\circ\mathrm{C}$, $T_\mathrm{h}=8.7~^\circ\mathrm{C}$, and $T_\mathrm{i}=-1.2~^\circ\mathrm{C}$. Red dashed contours outline the pocket. The scale bar is $0.1~\mathrm{mm}$. (e) Temporal evolution of the corresponding pocket in (d): (e1) pocket length $l$ and width $w$, (e2) the tip velocities $v_\mathrm{f}$ and $v_\mathrm{r}$, defined as positive toward the interface. The vertical dashed line marks the onset of front-tip deceleration as the pocket approaches the interface, whereas the vertical dash-dotted line marks the transition from pocket elongation to shortening.}
\label{fig1}
\end{figure*}

Freezing and melting in multicomponent materials involve the coupled transport of heat, solute, and fluid, which collectively govern phase-boundary motion and the redistribution of liquid within the solid \cite{huerre2025freezing,du2024freezingflows,cenedese2023icebergs,wang2025icingreview,deville2022complex}. This coupling underlies a broad range of processes, from the evolution of natural ice to the formation of porous materials and the redistribution of melt in geological and synthetic solids \cite{wang2021icegrowth,du2023seawatermodes,wegst2024freeze,wang2024selflifting,schianoTranscrystallineMeltMigration2006,songLocalisedStructuringMetalsemiconductor2022,chu2024interfacial}. Natural sea ice exhibits spatially heterogeneous brine microstructures, with variations in brine volume fraction and pore connectivity across different depths \cite{goldenPercolationPhaseTransition1998}, as shown in Fig.~\ref{fig1}(a) \cite{golden2026percolation}. Beyond this spatial heterogeneity, an important question is how liquid retained within the solid migrates, deforms, and eventually escapes as the surrounding phase structure evolves. Understanding this behavior requires elucidating not only how heat and mass transfer drive liquid redistribution, but also how evolving phase boundaries modify the transport pathways that sustain it.

A well-known mechanism for liquid-inclusion migration is temperature-gradient zone melting (TGZM), also known as thermomigration. Local phase equilibrium imposes different solute concentrations at the warm and cold ends of an inclusion. The resulting solute redistribution couples melting at the warm end and freezing at the cold end, translating the inclusion through the solid \cite{tillerMigrationLiquidZone1963,anthonyThermalMigrationLiquid1971,hoekstraMigrationLiquidInclusions1965}. Classical theories and experiments established the roles of the imposed thermal gradient, crystallographic orientation, inclusion size, and interfacial kinetics \cite{clineEffectsMagnitudeCrystallographic1972,olanderThermalGradientMigration1982}. More recent models have addressed nonlinear migration and instability \cite{pawarAnalysisTemperatureGradient2022,kraitzmanSlowMigrationBrine2022}, as well as the coupling between inclusion shape and crystallization kinetics \cite{garmashovModelShapesLiquid2021,garmashovEvolutionCrosssectionalShapes2024}. These studies provide a framework for migration within the bulk solid, where inclusion motion is governed by the local thermal and solutal fields.

A macroscopic solid--liquid boundary changes this transport environment. For objects encountered by advancing freezing fronts, solute redistribution modifies trapping and deformation \cite{dedovetsFiveDimensional2018,tyagiThermalSolute2020,tyagiSoluteEffects2022}, while flow in intervening liquid films deforms engulfed droplets \cite{meijerThinFilm2023}. Flow can also reshape the freezing front itself, through bulk convection \cite{wang2021icefrontshaping} or thermocapillary stresses around droplets \cite{vanBuurenDeformingIce2024}. Freezing can reorganize the topology of compound droplets and expel their internal liquid \cite{meijerTopological2024}. During thawing, retreating fronts release engulfed droplets and particles \cite{raffiFreezeThaw2026}, and bubbles emerging from melting ice modify flow and heat transfer \cite{wengrove2023bubbleMelting}. For brine inclusions, early experiments documented migration toward and puncture of an ice--water interface \cite{harrisonMeasurementBrineDroplet1965}, and subsequent theory considered finite-size effects during interface crossing \cite{shreveMigration1967}. However, how a nearby phase boundary modifies front-end melting and rear-end freezing, and thereby governs the transition from migration to release, remains poorly understood, particularly the respective roles of thermal and solutal transport.

Here, we investigate the migration and release of brine pockets in saline ice using a quasi-two-dimensional Hele-Shaw cell, which allows us to track individual pockets, resolving the local velocities and deformation at the front (warm) and rear (cold) ends. During bulk migration, the pockets elongate as the front end advances faster than the rear, owing to the different temperature-dependent migration mobilities at the two ends. As they approach the ice--water boundary, the front end decelerates because the nearby boundary suppresses the local enhancement of the thermal gradient and thus reduces the thermal driving for melting. Upon liquid connection, the rear end accelerates as the opening provides an additional pathway for salt exchange with the surrounding liquid, enhancing rear-end freezing. These opposing responses drive a transition from elongation to rapid shortening. We further establish a quantitative description of the release dynamics by considering a diffusion-resistance model that accounts for both solute transport within the embedded pocket and mass exchange with the external liquid. Our findings reveal how a macroscopic phase boundary regulates the migration, deformation, and release of liquid inclusions through distinct but coupled changes in heat and solute transport pathways.

\section{Results}
\subsection{Migration and deformation of individual pockets}
A $200\,\mu\mathrm{m}$-thick Hele--Shaw cell filled with a $0.1~\mathrm{wt}\%$ aqueous NaCl solution [Fig.~\ref{fig1}(b,c)] is placed on two independently controlled copper blocks, to impose a horizontal temperature gradient across the sample. After freezing, the macroscopic ice--water boundary becomes quasi-stationary, while liquid pockets embedded in the ice continue to migrate. The effective sample-plane temperatures at the cold and hot sides are estimated as $T_\mathrm{c}=-3.3~^\circ\mathrm{C}$ and $T_\mathrm{h}=8.7~^\circ\mathrm{C}$, respectively. A steady heat-flux balance, using the measured mean ice thickness $L$, gives an ice--water interfacial temperature of $T_\mathrm{i}=-1.2~^\circ\mathrm{C}$ and a mean thermal gradient in the ice of $G_\mathrm{s}=(T_\mathrm{i}-T_\mathrm{c})/L$. To characterize the non-rigid migration, we track the two ends of each pocket relative to the instantaneous local ice--water interface, as illustrated in Fig.~\ref{fig1}(c). The end closer to the boundary is defined as the front, and the opposite end as the rear. Their respective distances from the interface are denoted by $d_\mathrm{f}$ and $d_\mathrm{r}$, giving the pocket length $l=d_\mathrm{r}-d_\mathrm{f}$. The corresponding tip velocities are $v_\mathrm{f}=-\dot d_\mathrm{f}$ and $v_\mathrm{r}=-\dot d_\mathrm{r}$, with positive velocities directed toward the macroscopic boundary. Under the quasi-stationary interfacial condition, the advance of the front and rear tips corresponds to melting and freezing, respectively. Experimental procedures, thermal calibration, and trajectory extraction are described in Methods.

Fig.~\ref{fig1}(d) shows a full-field view of a representative brine pocket migrating toward the ice–water boundary. The dashed rectangle indicates the region shown in the time-lapse images. The image sequence in Fig.~\ref{fig1}(d) shows a representative brine pocket migrating toward the ice--water boundary while elongating along the imposed temperature gradient. The width at the midpoint of the pocket varies relatively little, whereas its length increases substantially [Fig.~\ref{fig1}(e1)]. Tracking the two ends separately reveals the origin of this deformation [Fig.~\ref{fig1}(e2)]. During bulk migration, before the vertical dashed line, both tips advance toward the boundary, with the front end moving faster than the rear end. Consequently, the length-change rate, $\dot l=v_\mathrm{f}-v_\mathrm{r}$, remains positive, accounting for the progressive elongation of the pocket.

The dynamics change as the pocket approaches and connects to the macroscopic ice--water boundary. Between the vertical dashed and dash-dotted lines, the front velocity decreases from its maximum, indicating slower front-end melting. Near the onset of liquid connection, the rear velocity increases markedly, indicating accelerated rear freezing. The rear velocity then overtakes the front velocity, marking the transition from pocket elongation to shortening, which is marked by the dash-dotted line [Fig.~\ref{fig1}(e)]. Similar sequences occur across the tracked pockets (See Methods for more information). 

The velocity data in Fig.~\ref{fig1}(e2) include all experimentally resolvable data points extracted from the tracked trajectories. As the pocket approaches and connects to the ice--water boundary, the front-tip velocity decreases and eventually becomes comparable to the rear-tip velocity, corresponding to the attainment of the maximum pocket length. Although liquid connection and maximum pocket length are physically distinct events, they occur in close temporal proximity in our experiments. Moreover, the onset of liquid connection is difficult to identify precisely from the images. We therefore use the sign of the length-change rate, $\dot l=v_\mathrm{f}-v_\mathrm{r}$, as a practical indicator of the pocket's connection state later, with positive and negative values characterizing the elongating and shortening regimes, respectively.

\subsection{Temperature-gradient-driven migration in the bulk}
We first consider the pocket migration far from the macroscopic interface, where its influence on the local transport fields is negligible. Since thermal diffusion is much faster compared with solute redistribution and pocket motion, the temperature field could be treated as quasi-steady \cite{anderson2020mushy}. Under local liquidus equilibrium, the interfacial temperature determines the equilibrium brine salinity \cite{hoekstraMigrationLiquidInclusions1965,lightHydrohaliteColdSea2009}. Because the rear is colder than the front, its equilibrium salinity is higher. This salinity difference drives salt diffusion toward the warmer front, coupling melting at the front with freezing at the rear and thereby translating the liquid pocket toward higher temperatures \cite{du2025seaiceaging,wang2026freezingaging}.

The migration velocity is determined by the local thermal gradient and the temperature-dependent solute transport at each end. The pocket ends are curved [Fig.~\ref{fig1}(c)], so the conductivity contrast between ice and brine modifies the local thermal field. To account for this effect, we approximate each tip locally as a circular interface and incorporate the resulting thermal-gradient enhancement into the migration law. Combining interfacial salt conservation with the liquid--solid density difference \cite{hoekstraMigrationLiquidInclusions1965,tillerMigrationLiquidZone1963,seidenstickerCommentPaperHoekstra1966a,du2025seaiceaging}, the migration velocity of either end can be expressed as
\begin{equation}
v_j=\chi_\rho\chi_{G,j}M(T_j)G_\mathrm{s},
\label{eq1}
\end{equation}
where $\chi_\rho=\rho_\mathrm{l}/\rho_\mathrm{s}\approx1.09$ is the liquid-to-solid density ratio, with $\rho_\mathrm{l}$ and $\rho_\mathrm{s}$ denoting the densities of liquid brine and ice, respectively. The factor $\chi_{G,j}=G_j/G_\mathrm{s}$ accounts for the local thermal-gradient enhancement at tip $j$, where $G_j$ is the magnitude of the temperature gradient on the liquid side of the tip and $G_\mathrm{s}$ is the mean temperature gradient in the surrounding ice; $j=\mathrm{f},\mathrm{r}$ denotes the front and rear tips, respectively. The temperature-dependent migration mobility is $M(T_j)=-[D(T_j)/S(T_j)]\cdot(\text{d}S/\text{d}T)|_{T_j}$, where $D(T_j)$ is the solute diffusivity in the liquid at the local tip temperature $T_j$, and $S(T_j)$ is the corresponding equilibrium brine salinity given by the liquidus relation.

For a circular liquid inclusion in an infinite solid subjected to a uniform far-field temperature gradient $G_\mathrm{s}$, the two-dimensional conduction solution gives a uniform liquid-side gradient with
\begin{equation}
\chi_{G,j}=\frac{G_j}{G_\mathrm{s}}
=\frac{2k_\mathrm{s}}{k_\mathrm{s}+k_\mathrm{l}}
\approx 1.57,
\qquad j=\mathrm{f},\mathrm{r},
\label{eq2}
\end{equation}
where $k_\mathrm{s}$ and $k_\mathrm{l}$ are the thermal conductivities of ice and liquid brine, respectively \cite{tillerMigrationLiquidZone1963}.

\begin{figure*}[t]
\centering
\includegraphics[width=1\linewidth]{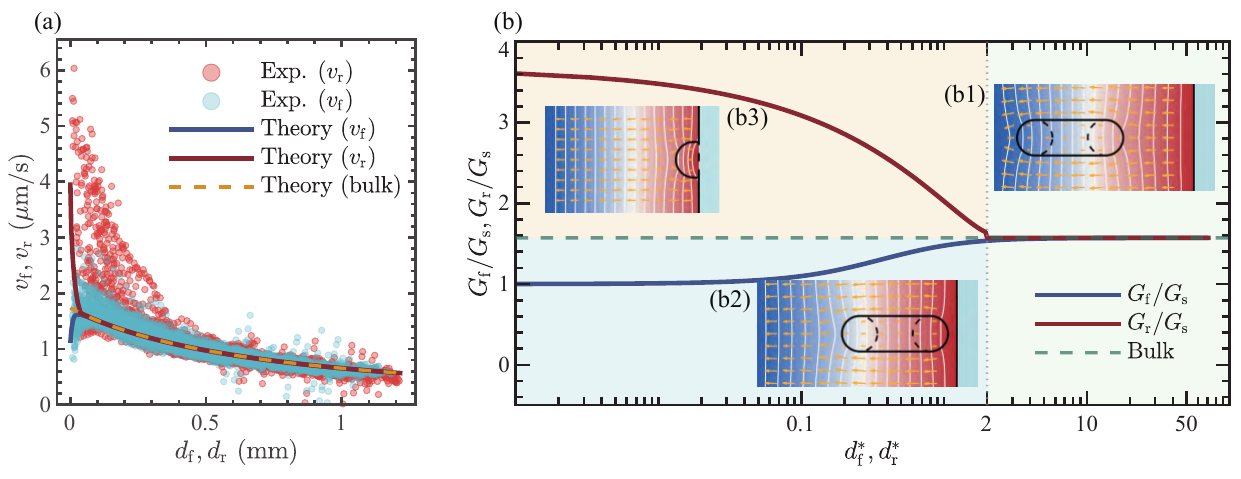}
\caption{Near-boundary thermal effects on front melting and rear freezing.
(a) Front and rear velocities versus their respective distances from the local ice--water interface, under the conditions of Fig.~\ref{fig1}. Cyan and red symbols denote front and rear measurements, respectively. The orange dashed curve is the bulk TGZM prediction, including the temperature dependence of the migration mobility. The blue and red solid curves include the front- and rear-end thermal boundary corrections, respectively, evaluated using the measured median width $w=33\,\mu\mathrm{m}$.
(b) Normalized liquid-side thermal gradients, $G_\mathrm{f}/G_\mathrm{s}$ and $G_\mathrm{r}/G_\mathrm{s}$, versus the corresponding normalized distances $d_\mathrm{f}^{*}=2d_\mathrm{f}/w$ and $d_\mathrm{r}^{*}=2d_\mathrm{r}/w$. The green dashed line denotes the position-independent bulk enhancement, $2/(1+K)$, where $K=k_\mathrm{l}/k_\mathrm{s}$. As the respective tip approaches the macroscopic interface, the front gradient decreases below the bulk value, whereas the connected rear gradient increases sharply only when the rear--interface distance becomes comparable to the pocket width. Insets show (b1) a pocket in the bulk, (b2) the front approaching the interface, and (b3) the connected rear geometry. Background colors, white curves, and orange arrows denote temperature, isotherms, and heat flux, respectively.}
\label{fig2}
\end{figure*}

In the bulk, the thermal-gradient enhancement is identical at the two tips, whereas the migration mobility varies with the local temperature. Consequently, the front and rear tips, which occupy different positions along the imposed temperature gradient at a given time, generally move at different velocities, as shown in Fig.~\ref{fig1}(e). When compared at the same distance from the macroscopic interface, the measured front- and rear-tip velocities collapse onto a common trend [Fig.~\ref{fig2}(a)], consistent with their same local thermal and solutal conditions at the same place. Combining Eq.~\eqref{eq1} and Eq.~\eqref{eq2} yields the bulk TGZM prediction (orange dashed curve in Fig.~\ref{fig2}(a)), capturing the position-dependent trend of the front and rear velocities. As the pockets approach the macroscopic boundary, however, the front- and rear-tip velocities exhibit distinct deviations from the bulk prediction: the front velocity decreases, whereas the rear velocity increases. The emergence of these opposing trends indicates that the macroscopic interface introduces additional effects beyond the bulk TGZM mechanism. We therefore next examine how the nearby ice--water boundary modifies the local thermal field and suppresses front-end melting.

\subsection{Near-boundary thermal effects on pocket migration}
To quantify how the nearby ice--water boundary modifies the thermal field, we approximate the front cap of a brine pocket as a circular liquid inclusion of radius $a=w/2$, separated by a gap $d_\mathrm{f}$ from a planar ice--water boundary [Fig.~\ref{fig2}(b2)]. The planar boundary is held at $T_\mathrm{i}$, and the imposed temperature gradient is recovered far from the inclusion. We solve the resulting half-infinite two-material steady heat-conduction problem analytically in bipolar coordinates (Supplementary Material), obtaining the local thermal-gradient correction $\chi_{G,\mathrm{f}}$ as a function of the normalized gap $d_\mathrm{f}^{*}=d_\mathrm{f}/a=2d_\mathrm{f}/w$ and the conductivity ratio $K=k_\mathrm{l}/k_\mathrm{s}$. This normalization identifies the pocket width as the characteristic length scale of the boundary effect. We use the measured median width, $w=33\,\mu\mathrm{m}$, to evaluate the correction and calculate the front velocity in Fig.~\ref{fig2}(a).

For fixed conductivities, the classical infinite bulk solution gives a position-independent gradient enhancement of $2/(1+K)$ at both ends [green dashed line in Fig.~\ref{fig2}(b)]. Near the macroscopic boundary, in contrast, the front gradient departs from this constant value. As $d_\mathrm{f}^{*}$ decreases, the isothermal boundary increasingly constrains the distortion of the isotherms around the front cap, causing $\chi_{G,\mathrm{f}}$ to decrease toward unity [blue curve in Fig.~\ref{fig2}(b)]. The local gradient enhancement is therefore progressively suppressed as the front approaches the ice-water boundary. This effect competes with the temperature dependence of the migration mobility: the increasing $M(T_\mathrm{f})$ favors faster migration toward the warmer boundary, whereas the decreasing $\chi_{G,\mathrm{f}}$ reduces the thermal driving. Incorporating this competition into Eq.~\eqref{eq1} gives
\begin{equation}
v_\mathrm{f}
=
\chi_\rho\chi_{G,\mathrm{f}}(d_\mathrm{f}^{*};K)
M(T_\mathrm{f})G_\mathrm{s}.
\label{eq3}
\end{equation}

This competition produces a nonmonotonic velocity response: the front velocity initially increases, reaches a maximum, and then decreases near the interface. The model curve given by Eq.~\eqref{eq3} well captures the measured front-end deceleration [blue curve in Fig.~\ref{fig2}(a)], linking the slower front melting to suppression of the local thermal-gradient enhancement.

We next assess the thermal contribution to rear-end acceleration after liquid connection. The connected pocket forms a liquid finger linking the rear cap to the external liquid [Fig.~\ref{fig2}(b3)], changing the conduction geometry. We determine the corresponding correction $\chi_{G,\mathrm{r}}$ as a function of the normalized rear-tip distance $d_\mathrm{r}^{*}=d_\mathrm{r}/a=2d_\mathrm{r}/w$. When the planar ice--water boundary intersects the circular rear cap, the correction can be obtained analytically. For the earlier connected stage, when a finite-width liquid finger separates the rear cap from the external liquid, we solve the conduction problem numerically (Supplementary Material). Numerical calculations show that the rear thermal gradient is appreciably affected by the macroscopic boundary only when $d_\mathrm{r}$ becomes comparable to the pocket width $w$ (Supplementary Material). We therefore combine the conventional infinite-domain solution with a near-boundary analytical approximation based on a circular cap intersected by the boundary. The slight discontinuity at the transition between the two solutions [Fig.~\ref{fig2}(b)] arises from the piecewise approximation.

The resulting rear gradient shows a trend opposite to that at the front. It remains close to the bulk value while the rear is far from the macroscopic ice--water boundary, but increases strongly as $d_\mathrm{r}$ becomes comparable to the pocket width [red curve in Fig.~\ref{fig2}(b)]. It is noteworthy that, in this model, liquid connection does not immediately produce a substantial thermal correction at the rear; the strong gradient enhancement develops only when the rear itself approaches the interface. Using the same representative width, the thermal model gives
\begin{equation}
v_\mathrm{r}^{\mathrm{th}}
=
\chi_\rho\chi_{G,\mathrm{r}}(d_\mathrm{r}^{*};K)
M(T_\mathrm{r})G_\mathrm{s}.
\end{equation}

The thermal model does predict an upturn in the rear velocity after connection [red curve in Fig.~\ref{fig2}(a)]. However, pronounced acceleration begins when the rear end becomes much closer to the macroscopic boundary than that observed in experiments. Moreover, the predicted velocity rises more sharply with decreasing $d_\mathrm{r}$ than the experimental trend. The model therefore captures the qualitative tendency toward faster rear freezing, but fails to reproduce both the spatial onset and the more gradual development of the observed acceleration. These discrepancies indicate that thermal-field modification alone is insufficient to explain the post-connection rear dynamics.

\subsection{Connection-induced acceleration of rear-end freezing}

\begin{figure*}[t]
\centering
\includegraphics[width=1\linewidth]{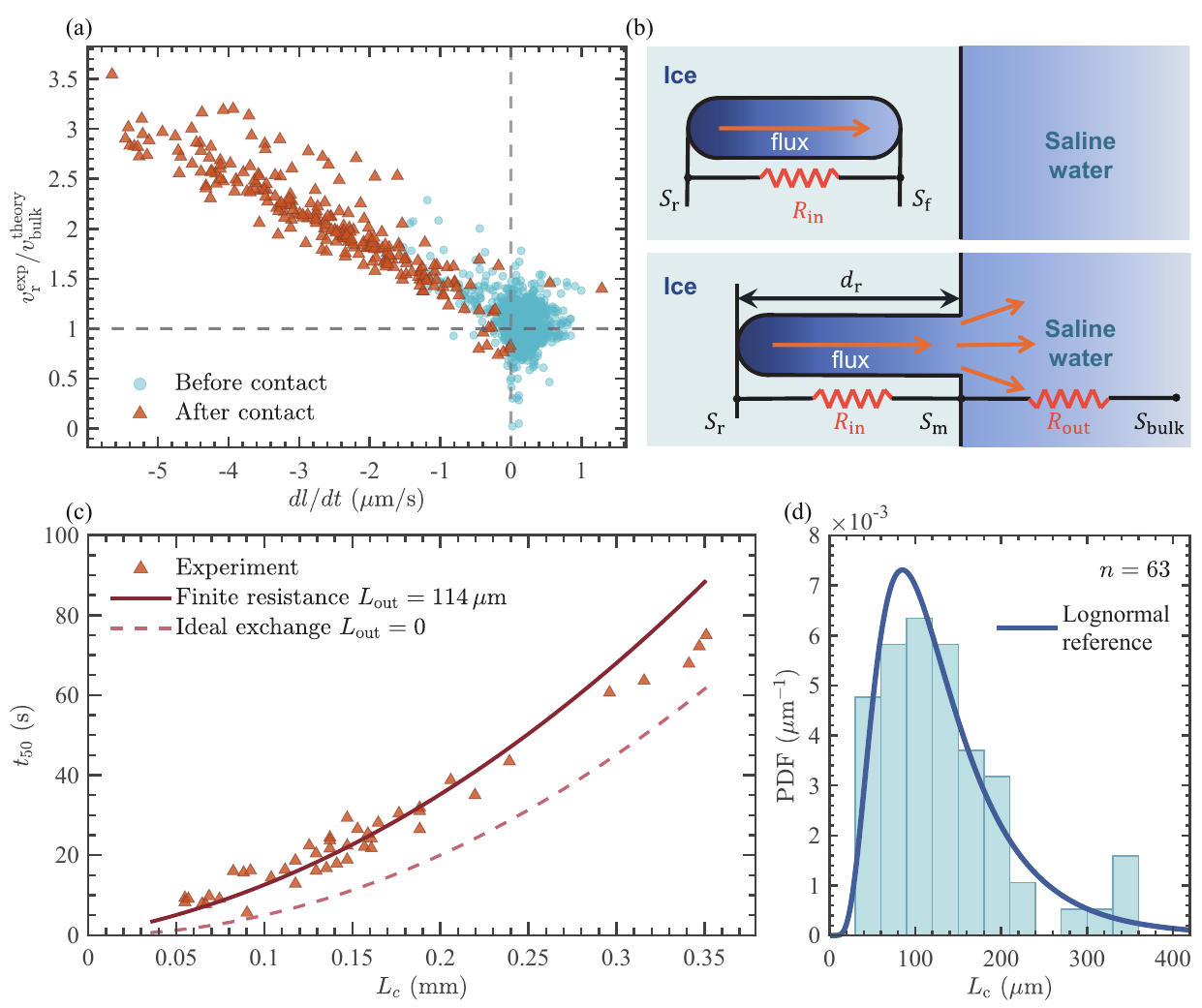}
\caption{Liquid connection promotes rear freezing through external salt exchange.
(a) Measured rear velocity normalized by the bulk TGZM prediction versus the pocket length-change rate. Cyan circles and orange triangles denote measurements before and after optically identified liquid connection, respectively. The dashed lines mark zero length-change rate and unit normalized velocity.
(b) Schematics of salt transport before connection (top) and after connection (bottom). $S_\mathrm{r}$ and $S_\mathrm{f}$ denote the equilibrium salinities at the rear and front tips, $S_\mathrm{m}$ is the opening salinity, and $S_\mathrm{bulk}$ is the background salinity of the external liquid. $R_\mathrm{in}$ and $R_\mathrm{out}$ represent diffusion resistance within the embedded pocket and the effective resistance during exchange across the external salinity boundary layer, respectively. Orange arrows indicate salt transport.
(c) Half-length release time $t_{50}$ versus the embedded length at connection, $L_c=d_\mathrm{r}(t_c)$, where $t_c$ is determined from the images. The half-length release time is the interval from $t_c$ to the first crossing of $d_\mathrm{r}=L_c/2$. Triangles denote events with a resolved half-length crossing. The solid curve is the diffusion-resistance model with a shared fitted exchange length $L_\mathrm{out}=114\,\mu\mathrm{m}$; the dashed curve is the ideal-exchange limit, $L_\mathrm{out}=0$. Both curves retain the temperature dependence of the diffusivity and equilibrium salinity. (d) Probability density of the connection length $L_c$ for the 63 retained pockets. The blue curve is a lognormal reference.}
\label{fig3}
\end{figure*}

Figure~\ref{fig3}(a) shows the measured rear velocity, normalized by the bulk TGZM prediction given by Eq.~\eqref{eq1} and Eq.~\eqref{eq2}, as a function of the pocket length-change rate $\dot l=dl/dt$. Measurements in the elongating regime, $\dot l>0$, cluster near unity, whereas those in the shortening regime, $\dot l<0$, show pronounced rear-velocity enhancement. Rear acceleration thus accompanies the transition from elongation to shortening. 
As shown in Fig.~\ref{fig1}(e) and Section II.A, this transition occurs close to the optically identified onset of liquid connection with the external liquid.
We now examine how liquid connection changes the salt transport sustaining rear-end freezing. 

Liquid connection changes the solute-transport boundary condition [Fig.~\ref{fig3}(b)]. In an isolated pocket, salt is redistributed between the freezing rear and melting front. After connection, salt rejected by rear freezing can diffuse through the opening into the more dilute external liquid. The thermal field continues to determine the rear temperature and its equilibrium salinity, while the salt flux depends on transport through both the remaining pocket and the salt-enriched liquid adjacent to the macroscopic ice--water boundary.

We consider quasi-steady diffusion and represent transport within the pocket and across the external salinity boundary layer by two resistances in series. For a liquid finger of approximately uniform cross section, we use the rear-end diffusivity $D_\mathrm{r}=D(T_\mathrm{r})$ as the representative diffusivity along the pocket, giving $R_\mathrm{in}=d_\mathrm{r}/D_\mathrm{r}$. The external resistance $R_\mathrm{out}$ accounts for the salinity drop between the opening and the outer edge of the boundary layer, where the salinity is $S_\mathrm{bulk}$. The outward salt mass flux per unit pocket cross-sectional area is then $J=\rho_\mathrm{l}(S_\mathrm{r}-S_\mathrm{bulk})/(R_\mathrm{in}+R_\mathrm{out})$, where $S_\mathrm{r}=S(T_\mathrm{r})$. Combining this flux with local solute conservation at the freezing rear, $J=\rho_\mathrm{s}v_\mathrm{r}S_\mathrm{r}$, gives
\begin{equation}
v_\mathrm{r}=
\frac{\chi_\rho D_\mathrm{r}}{d_\mathrm{r}+L_\mathrm{out}}
\left(1-\frac{S_\mathrm{bulk}}{S_\mathrm{r}}\right),
\label{eq:release}
\end{equation}
where $L_\mathrm{out}=D_\mathrm{r}R_\mathrm{out}$ is the effective external exchange length. As rear freezing shortens the internal diffusion path, its resistance decreases and the rear velocity increases. We use a shared effective exchange length to represent the additional resistance of the external salinity boundary layer.

Connected pockets span a broad range of lengths and shapes, and their final disappearance is difficult to identify consistently. We therefore characterize release by the half-length release time $t_{50}$. For each pocket, $L_c=d_\mathrm{r}(t_c)$ is the embedded length at the optically identified connection time $t_c$. We define $t_{50}$ as the interval from $t_c$ to the first crossing of $d_\mathrm{r}=L_c/2$, determined directly from the rear-position trajectory. This measure compares pockets at the same fractional reduction of their initial embedded length. Evaluating $D_\mathrm{r}$ and $S_\mathrm{r}$ at connection and integrating $\dot d_\mathrm{r}=-v_\mathrm{r}$ from $d_\mathrm{r}=L_c$ to $L_c/2$ gives
\begin{equation}
t_{50}=
\frac{1}
{\chi_\rho D_\mathrm{r}\left(1-S_\mathrm{bulk}/S_\mathrm{r}\right)}
\left(\frac{3L_c^2}{8}+\frac{L_\mathrm{out}L_c}{2}\right).
\label{eq:half_length_time}
\end{equation}
The two resistances contribute different length dependences to the release time: a quadratic contribution from internal diffusion and a linear contribution from exchange across the external boundary layer.

For comparison with the measured half-length release times in Fig.~\ref{fig3}(c), we integrate Eq.~\eqref{eq:release} while retaining the temperature dependence of $D_\mathrm{r}$ and $S_\mathrm{r}$ (Methods). Setting $L_\mathrm{out}=0$ gives the ideal-exchange limit, in which the opening salinity equals the external background value. This prediction systematically underestimates the measured times [dashed curve in Fig.~\ref{fig3}(c)]. The additional time associated with external resistance is expressed by the term proportional to $L_\mathrm{out}L_c$ in Eq.~\eqref{eq:half_length_time}. Fitting a single exchange length across the events with resolved half-length crossings yields $L_\mathrm{out}=114\,\mu\mathrm{m}$, which captures the overall dependence of $t_{50}$ on $L_c$ [solid curve in Fig.~\ref{fig3}(c)].

Salt conservation provides an independent estimate of this length scale. For the initial salinity $S_0=0.1~\mathrm{wt}\%$ and measured ice-layer thickness $L$, we approximate the external salinity profile as a linear decrease from the interfacial equilibrium value $S_\mathrm{i}\approx2~\mathrm{wt}\%$ to $S_0$. Equating the salt rejected during ice formation to the excess salt stored in this layer gives $\delta\simeq2S_0L/(S_\mathrm{i}-S_0)\approx130\,\mu\mathrm{m}$. For comparable internal and external diffusivities, $L_\mathrm{out}\approx\delta$, in reasonable agreement with the fitted $114\,\mu\mathrm{m}$. Salt retained in the embedded pockets reduces the inventory available for external enrichment and hence the layer thickness, consistent with the fitted length being smaller than the simplified salt-balance estimate.

The internal and external contributions to $t_{50}$ are equal at $L_c=4L_\mathrm{out}/3\approx152\,\mu\mathrm{m}$. External exchange makes a larger relative contribution for shorter pockets, while internal diffusion becomes increasingly important for longer pockets. The measured connection lengths span both sides of this crossover [Fig.~\ref{fig3}(d)]. We describe their distribution with a lognormal reference, a form also used for brine-inclusion sizes in sea ice \cite{perovich1996quantitative}. Liquid connection thus couples rear freezing to the external salinity field, and the release timescale is governed jointly by the remaining internal diffusion length and the resistance to external exchange.

\section{Discussion}
In summary, we have identified how thermal proximity and liquid connectivity regulate the distinct phase-change processes driving brine-pocket migration and release. Suppression of the local thermal-gradient enhancement slows front melting, while connection to the external liquid accelerates rear freezing through solute exchange. Together, these responses convert bulk elongation into rapid shortening. A quasi-steady diffusion-resistance model relates the characteristic release time to the pocket length at connection and the external resistance to mass transfer.

These findings provide a pore-scale link between inclusion migration and diffusion-driven desalination in aging saline ice \cite{du2025seaiceaging}. Release connects liquid redistribution within the ice to salt exchange with the surrounding liquid. Larger-scale descriptions must also account for liquid-network connectivity and convective transport \cite{notzDesalination2009,middletonRelativeRoleShort2022}. These processes couple desalination to changes in porosity and heat transport, linking individual release events to the evolution of the ice layer \cite{wang2026freezingaging}.

More broadly, controlling temperature gradients and composition is central to tailoring microstructure during solidification \cite{meijerBubbleGrowth2024,thievenazBubbleShape2025,shao2025manipulating}. Our results show that liquid connectivity provides a further means of regulating mass exchange and the phase change it sustains. Coordinating thermal and compositional boundary conditions with these exchange pathways offers a physical basis for directing inclusion migration, retention, and release, linking local transport mechanisms to microstructural design \cite{deville2009metastable,deville2017freezing}.

\section{Methods}
\subsection{Experimental setup}
The experiments were conducted in a Hele-Shaw freezing cell with a gap thickness of $200~\mu \mathrm{m}$ [Fig.~1(b,c)]. Before each experiment, the cell was filled with a $0.1~\mathrm{wt}\%$ aqueous NaCl solution. Then, two ends of the cell were placed in thermal contact with two independently controlled copper blocks (separated by approximately $H=2.8~\mathrm{mm}$). The temperatures of the copper blocks were controlled using Peltier elements under PID feedback. The initial cold- and hot-side copper-block temperature setpoints were $T^\mathrm{Cu}_\mathrm{c}=-4.0 ~^\circ\mathrm{C}$ and $T^\mathrm{Cu}_\mathrm{h}=8.0 ~^\circ\mathrm{C}$, respectively. These temperatures were maintained for $20~\mathrm{min}$ to establish a steady temperature field. Afterwards, a small ice seed was artificially introduced at the cold end to initiate freezing. After the advancing ice-water interface became stationary, the cold-side copper temperature $T^\mathrm{Cu}_\mathrm{c}$ was decreased from $-4.0$ to $-10.0~^{\circ}\mathrm{C}$ at a constant rate of $0.01~\mathrm{K/s}$, while the hot-side setpoint $T^\mathrm{Cu}_\mathrm{h}$ was maintained at $8.0~^{\circ}\mathrm{C}$. Both setpoints were held constant after the cooling ramp. The sidewalls of the cell and the exposed surfaces of the copper blocks were covered with thermally insulating cotton, except for the optical access required for imaging. The evolution of the ice-brine microstructure was recorded in a top-view configuration using a camera equipped with a long-working-distance lens, with a frequency of 1 frame per second. The sample was illuminated with a diffuse LED source to minimize local heating of the sample. Dry nitrogen was supplied to the chamber to suppress condensation, while the chamber humidity was monitored throughout the experiment.

\begin{figure*}[!htb]
\centering
\includegraphics[width=1\linewidth]{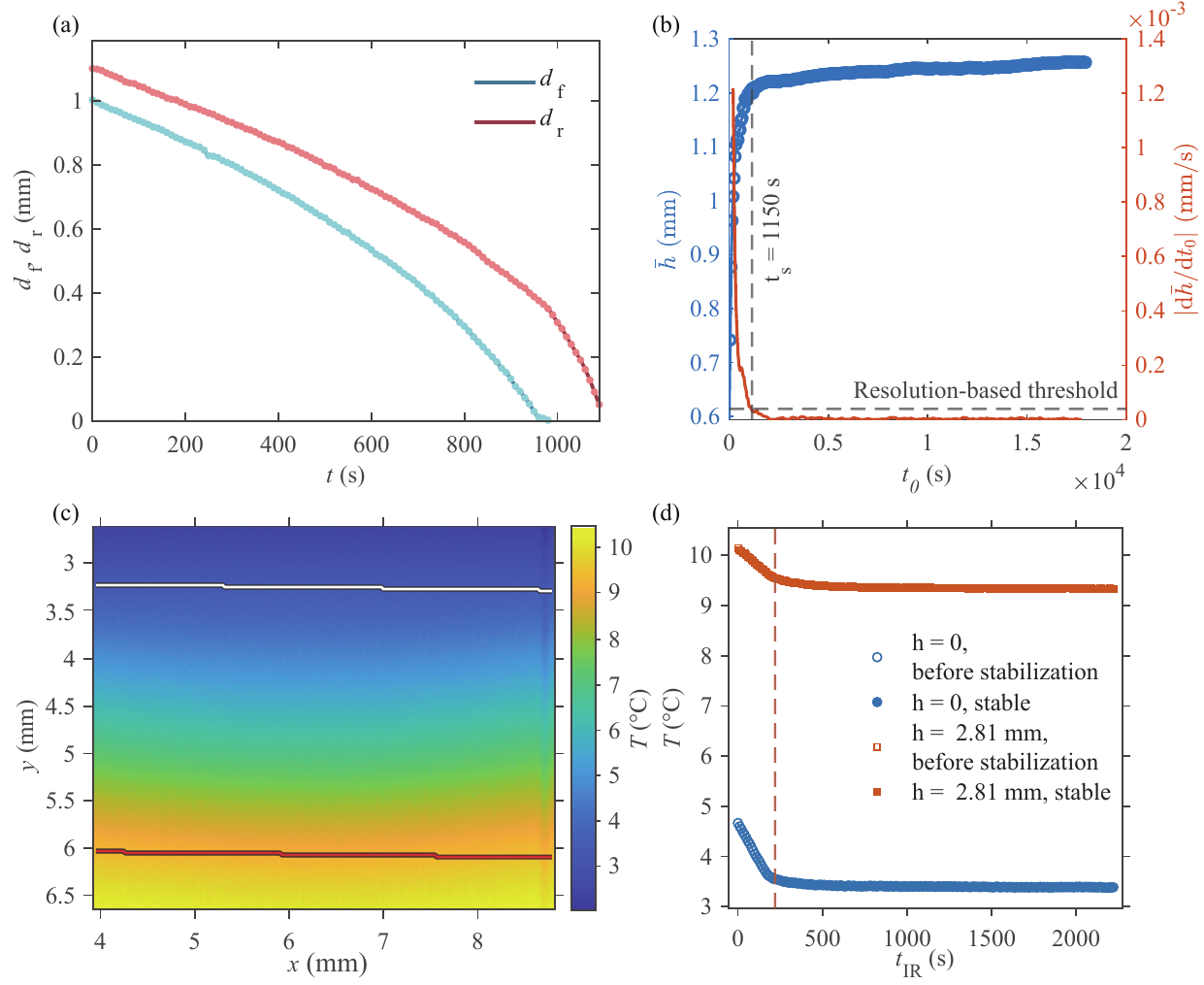}
\caption{Experimental characterization and measurements. (a) Representative temporal evolution of the distances $d_{\mathrm{f}}$ and $d_{\mathrm {r}}$ from the front and rear ends of a brine pocket to the ice–water interface (corresponding to Fig.~\ref{fig1}(e)). (b) Temporal evolution of the spatially averaged ice–water interface height $\bar{h}$ and its locally fitted rate of change $|d\bar{h}/dt_0|$. The horizontal dashed line indicates the resolution-based threshold, and the vertical dashed line marks the onset of the quasi-stationary stage at $t_\mathrm{s}=1150\,\mathrm{s}$. (c) Representative infrared image showing the temperature distribution on the upper glass surface of the Hele–Shaw cell. The white and red lines mark the cold and hot ends of the cell, respectively. (d) Temporal evolution of the upper-surface temperatures at the cold $h=0$ and hot $h=2.81\, \mathrm{mm}$ ends. Open and filled symbols distinguish measurements before and after the quasi-stationary stage, respectively. The optical measurements in (b) and the infrared measurements in (c,d) were performed in separate experiments under identical experimental conditions, accounting for the difference in the onset times of the quasi-stationary stage.
}
\label{fig4}
\end{figure*}

\subsection{Imaging process}
The brine pockets were segmented from the optical images using the SAM2 model implemented on the Supervisely platform, with manual assistance to ensure consistent pocket identification between frames. Although the original videos were recorded at $1~\mathrm{fps}$, pocket morphology and position were analyzed
every 10 frames, corresponding to a sampling interval of $\Delta t=10~\mathrm{s}$. The front and rear tips were defined as the two extrema of the segmented pocket contour along the imposed temperature-gradient direction. The corresponding
local ice-water interface positions were identified independently from the optical images using MATLAB. The front- and rear-interface distances, $d_\mathrm{f}$ and
$d_\mathrm{r}$, were evaluated relative to the local interface at the respective tip positions. The pocket length and width were defined as $l=d_\mathrm{r}-d_\mathrm{f}$ and the transverse extent of the segmented pocket at the midpoint
between the two tips, respectively.

The tip velocities were obtained from the measured $d_\mathrm{f}(t)$ and $d_\mathrm{r}(t)$ trajectories [Fig.~\ref{fig4}(a) shows the typical data corresponding to Fig.~\ref{fig1}(e)] using a seven-point, second-order Savitzky--Golay differentiation.
For the first and last three points of each trajectory, the derivative was evaluated from a one-sided quadratic least-squares fit over the same seven-point window. Only trajectories containing at least 30 sampled positions were retained, so that the velocity analysis was based on tracks extending well beyond the local seven-point differentiation window. Contact with the ice-water interface was
identified when the segmented pocket overlapped the detected interface within a $\pm2$-pixel tolerance. The front velocity was evaluated only before the first contact, whereas the rear velocity was retained over the full trajectory of the
same pocket. Notably, not all tracked inclusions could be followed continuously until their eventual connection with the ice-water interface. Some inclusions lost their resolvable
identity within the ice before reaching the interface. This may arise from out-of-plane motion within the finite $200~\mu\mathrm{m}$ cell gap, causing an
inclusion to move outside the effective depth of field or to overlap with other microstructures along the optical axis. Because their subsequent trajectories could not be determined reliably, these inclusions were not included in the quantitative comparison with theory. Accordingly, the analysis presented in the main text was restricted to inclusions with sufficiently long trajectories and a clearly identified final
contact with the ice-water interface. Of the 166 segmented inclusions, $63$ satisfied these criteria and were retained for the theoretical analysis.

For the interface analysis, the mean interface position was evaluated at intervals of $\delta t=50~\mathrm{s}$. At each sampled time $t_0$, the local ice--water interface position $h(x,t_0)$ was spatially averaged along the transverse direction to obtain the mean interface height $\bar h(t_0)=\langle h(x,t_0)\rangle_x$.
The local migration rate $\mathrm{d}\bar h/\mathrm{d}t_0$ was estimated from a linear least-squares fit to nine consecutive measurements, corresponding to a $400~\mathrm{s}$ time window. The spatial resolution of the optical system is $\delta h=1.96\times10^{-3}~\mathrm{mm}$. We therefore define the onset of the quasi-stationary stage, $t_s$, by
\begin{equation}
\left|\frac{\mathrm{d}\bar h}{\mathrm{d}t_0}\right|
<
\frac{\delta h}{\delta t}
=
3.92\times10^{-5}~\mathrm{mm/s},
\label{S1}
\end{equation}
with this condition remaining satisfied for at least $250~\mathrm{s}$ thereafter [Fig.~\ref{fig4}(b)]. The time used in the main text is defined as $t=t_0-t_s$, so that $t=0$ corresponds to the onset of the quasi-stationary stage. The mean ice-layer thickness used in the thermal analysis is then $L=\left\langle \bar h(t_0)\right\rangle_{t_0\ge t_s}=1.24~\mathrm{mm}$.

\subsection{Measurement and estimation of interfacial temperature}

The temperatures $T_\mathrm{c}$, $T_\mathrm{h}$ reported in the main text denote the effective temperatures in the sample midplane and are distinct from the copper setpoints $T^\mathrm{Cu}_\mathrm{c}$ and $T^\mathrm{Cu}_\mathrm{h}$. As shown in Fig.~\ref{fig1}(b), at two ends of the cell, the lower glass plate was in thermal contact with the temperature-controlled copper blocks, whereas the upper glass plate remained exposed for optical access. The apparent temperature of the upper glass surface was measured using an infrared camera (FAST L200, Telops) through a germanium window (see a typical thermal image in Fig.~\ref{fig4}(c)). The complete optical path, including the glass surface and the germanium window, was calibrated against known reference temperatures measured by thermocouples under the same measurement configuration, yielding calibrated upper-glass-surface temperatures of $T^{\mathrm{upper}}_\mathrm{c}=3.4 ~^\circ\mathrm{C}$ and $T^{\mathrm{upper}}_\mathrm{h}=9.4 ~^\circ\mathrm{C}$ [Fig.~\ref{fig4}(d)]. At the cold and hot ends, the lower-surface temperatures were taken to be equal to the temperatures of the corresponding copper blocks. Assuming a linear temperature variation between the upper and lower glass surfaces, the effective sample-plane temperatures were estimated as $T_j=(T^\mathrm{upper}_j+T^\mathrm{Cu}_j)/{2}$, where $j=\mathrm{c},\mathrm{h}$. The resulting values $T_\mathrm{c}=-3.3~^\circ\mathrm{C}$, $T_\mathrm{h}=8.7~^\circ\mathrm{C}$, rather than the PID setpoints, were used in the subsequent analysis. 

The ice-water interfacial temperature $T_\mathrm{i}$ was inferred from a one-dimensional steady-state heat-conduction model.  During the quasi-stationary stage, the macroscopic temperature
field was treated as quasi-steady. Using the mean ice-layer thickness $L$ determined above, we considered an ice layer of thickness $L$ between the cold boundary and the mean ice-water interface, together with an external liquid layer of thickness $L_\mathrm{l}=H-L$ between the interface and the hot boundary. The corresponding boundary temperatures were the calibrated sample-plane temperatures $T_\mathrm{c}$ and $T_\mathrm{h}$. The thermal conductivities of ice and liquid water are denoted by $k_\mathrm{s}$ and $k_\mathrm{l}$, respectively, and the thermal conductivity of the dilute NaCl solution was approximated by that of pure water. Continuity of
the conductive heat flux across the mean ice-water interface then gives
\begin{equation}
T_\mathrm{i}=
\frac{(k_\mathrm{s}/L)T_\mathrm{c}+(k_\mathrm{l}/L_{\mathrm{l}})T_\mathrm{h}}
{k_\mathrm{s}/L+k_\mathrm{l}/L_{\mathrm{l}}}.
\label{S2}
\end{equation}

The calculated interfacial temperature $T_\mathrm{i}=-1.2 ~^\circ\mathrm{C}$ is below the equilibrium melting temperature of pure water. Physically, this subzero value is consistent with freezing-point depression caused by local salt enrichment near the ice-water interface. During the preceding formation of the ice layer, salt was rejected from the growing solid into the adjacent liquid. The strong confinement of the $200~\mu\mathrm{m}$ Hele-Shaw gap limits convective mixing, allowing the rejected salt to remain concentrated near the interface.

\subsection{Temperature-dependent migration mobility} 
The migration mobility $M(T_j)$ is determined from local phase equilibrium and solute diffusivity. For each end $j=\mathrm{f},\mathrm{r}$, the temperature $T_j$ is evaluated from the quasi-steady background temperature field at the instantaneous tip position. The corresponding equilibrium brine salinity $S_j$, expressed as a mass fraction, follows the NaCl liquidus relation \cite{10.2113/gsecongeo.83.1.197},
\begin{equation}
T_j=-60.37S_j-581.23S_j^3,    
\end{equation}
where temperature is in degrees Celsius. Accordingly, $(dS/dT)_{T_j}=-(60.37+1743.69S_j^2)^{-1}$. The effective NaCl diffusivity is approximated by the Nernst–Hartley relation for a binary 1:1 electrolyte \cite{ghaffariMDSimulationEvaluation2013},
\begin{equation}
D(T_j)=\frac{2D_{\mathrm{Na}^+}(T_j)D_{\mathrm{Cl}^-}(T_j)}
{D_{\mathrm{Na}^+}(T_j)+D_{\mathrm{Cl}^-}(T_j)}.
\end{equation}

The ionic diffusivities are evaluated using infinite-dilution reference values and temperature-dependent correlations \cite{madanisaniCriticalReviewModels2024}. These independently specified properties determine $M(T_j)=-[D(T_j)/S_j](dS/dT)_{T_j}$, while the geometric modification of the thermal field enters separately through the local gradient $G_j$.

\subsection{Effective model for post-connection transport} 
We represent the embedded liquid as a one-dimensional channel of approximately uniform cross section. The opening is located at $x=0$, and the freezing rear is at $x=d_\mathrm{r}$. The macroscopic ice–water interface is taken to be stationary during the shortening event. In the prescribed linear temperature field,
$T_\mathrm{r}=T_\mathrm{i}-G_\mathrm{s}d_\mathrm{r}$. The rear salinity and diffusivity are evaluated as $S_\mathrm{r}=S_{\mathrm{eq}}(T_\mathrm{r})$ and $D_\mathrm{r}=D(T_\mathrm{r})$ from the relationship above. Salinity is expressed as a mass fraction, and the liquid density $\rho_\mathrm{l}$ is treated as constant.

We consider steady, one-dimensional salt diffusion through an embedded pocket of uniform cross section, neglecting liquid advection. For a prescribed temperature field, the salinity satisfies
\begin{equation}
\frac{\mathrm d}{\mathrm dx}
\left[
D[T(x)]\frac{\mathrm dS}{\mathrm dx}
\right]=0.
\end{equation}

The outward salt mass flux therefore satisfies $J=\rho_{\mathrm{l}}(S_{\mathrm{r}}-S_\mathrm{m})/{R_{\mathrm{in}}(d_{\mathrm{r}})}$, and
\begin{equation}
R_{\mathrm{in}}(d_\mathrm{r})=
\int_0^{d_\mathrm{r}}\frac{\mathrm dx}{D[T(x)]}\approx d_\mathrm{r}/D_\mathrm{r}.
\end{equation}

The relative error of the rear-diffusivity approximation is quantified by
\begin{equation}
\epsilon_R(d_\mathrm{r})=
\left|
\frac{d_\mathrm{r}}{D_\mathrm{r}R_{\mathrm{in}}(d_\mathrm{r})}-1
\right|.
\end{equation}

For the plotted post-connection data, numerical evaluation of the integral gives a maximum relative error below $1.13\%$. Combining the exchange with the surrounding liquid, this relation eliminates the opening salinity and gives
\begin{equation}
J=\rho_\mathrm{l}
\frac{S_\mathrm{r}-S_{\mathrm{bulk}}}
{R_{\mathrm{in}}+R_{\mathrm{out}}}, 
\end{equation}
which yields Eq.~\eqref{eq:release} in the model above.


With a fixed opening, the embedded length $d_\mathrm{r}$ evolves according to $\dot d_\mathrm{r}=-v_\mathrm{r}(d_\mathrm{r})$. Integrating the velocity law gives
\begin{equation}
t_{50}(L_c;L_\mathrm{out})
=
\int_{L_c/2}^{L_c}
\frac{d_\mathrm{r}+L_\mathrm{out}}
{\chi_\rho D_\mathrm{r}(d_\mathrm{r})
\left[1-S_{\mathrm{bulk}}/S_\mathrm{r}(d_\mathrm{r})\right]}
\,\mathrm dd_\mathrm{r}.
\end{equation}

The mobility factor can be expressed as
\begin{equation}
C(d_\mathrm{r})=\chi_\rho D_\mathrm{r}(d_\mathrm{r})
\left[1-\frac{S_{\mathrm{bulk}}}{S_\mathrm{r}(d_\mathrm{r})}\right].
\end{equation}

It changes by less than $0.41\%$ during the rear freezing process, and can therefore be approximated by a constant $C_0=C(L_c)$ over the integration interval. In this limit, the integration can be simplified and yields Eq.~\eqref{eq:half_length_time} in the model above.

Meanwhile, it is noteworthy that half-length release times in our experiment were determined from the first observed crossing of $d_\mathrm{r} = L_c/2$, using linear interpolation between consecutive frames. Events without a resolved crossing were excluded from the scatter plot of half-length release times. Incomplete observations arose from limited post-contact tracking, loss of identifiable pocket contours, and, in some cases, coalescence with neighboring liquid pockets.

\vspace{20pt}
\noindent
{\bf Data Availability:} 
All relevant data that support the results of this study are presented in the main text and Supplementary Information.

\vspace{20pt}
\noindent
{\bf{Acknowledgement:}}
This work is supported by the National Natural Science Foundation of China Excellence Research Group Program for ‘Multiscale Problems in Nonlinear Mechanics’ (No.~12588201), the National Key R$\&$D Program of China under Grant No. 2025YFF0511800, the National Natural Science Foundation of China under Grant Nos. 12402321 and 12602337, the New Cornerstone Science Foundation through the New Cornerstone Investigator Program and the XPLORER PRIZE, the China Postdoctoral Science Foundation No. 2026M791982, and the Postdoctoral Fellowship Program of CPSF No. GZC20261056.

\vspace{20pt}
\noindent
{\bf{Author contributions:}}
X.T., F.W. and C.S. conceived the project. X.T., F.W., Y.D. and C.S. contributed to the study design. X.T. and Y.W. built the experimental apparatus and performed the experiments. X.T., F.W., Y.D. and J.H. analysed the data and carried out the theoretical modelling. X.T. and F.W. wrote the initial draft. All authors discussed the results and revised the manuscript. F.W. and C.S. supervised the research.

\vspace{20pt}
\noindent
{\bf{Competing Interests:}}
The authors declare no competing interests.

\setcounter{figure}{0} 
\setcounter{equation}{0} 
\renewcommand\thefigure{S\arabic{figure}}
\renewcommand\theequation{S\arabic{equation}}
\onecolumngrid

\section*{Supplementary Material}
\subsection*{Analytical front-end conduction solution}
\begin{figure*}[!htb]
\centering
\includegraphics[width=1\linewidth]{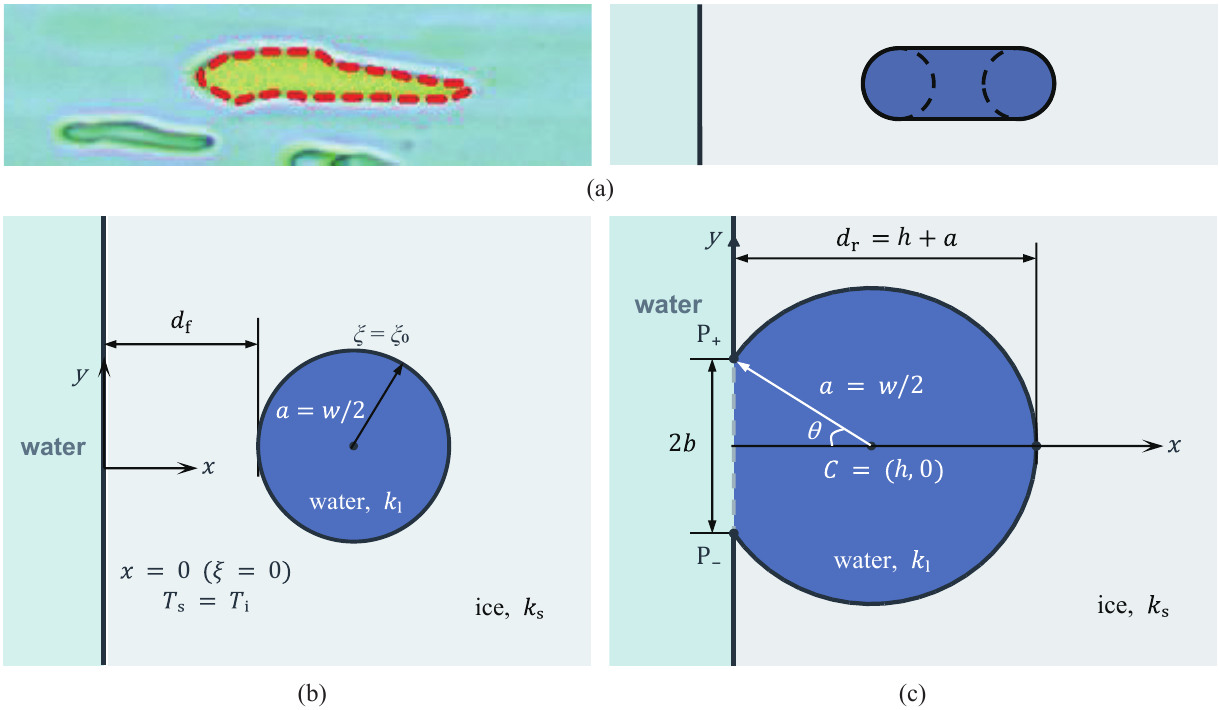}
\caption{Geometric approximations for the analytical conduction models. (a) Representative experimental image of an elongated brine inclusion (red dashed contour) and its idealized runway-shaped geometry, with semicircular end caps of radius $a=w/2$. (b) Front-end model before contact with the ice–water boundary. The front cap is approximated by a circular liquid inclusion separated from the planar isothermal boundary $T_\mathrm{s}=T_\mathrm{i}$ by a distance $d_\mathrm{f}$. In bipolar coordinates, the planar boundary and circular interface correspond to $\xi=0$ and $\xi=\xi_0$, respectively. (c) Rear-end model after connection to the external liquid. The local geometry is approximated by a circle centered at $C=(h,0)$, intersecting the planar boundary at $P_\pm=(0,\pm b)$, where $d_\mathrm{r}=h+a$, $h=a\cos\theta$, and $b=a\sin\theta$. The dashed segment represents the opening connecting the inclusion to the external liquid. Here, $k_\mathrm{s}$ and $k_\mathrm{l}$ denote the thermal conductivities of ice and water, respectively.
}
\label{figS2}
\end{figure*}

As shown in Fig.~\ref{figS2}(a), the observed liquid inclusions are idealized as runway-shaped regions, consisting of two semicircular end caps connected by straight sidewalls. The total transverse width is denoted by $w$, and the radius of each end cap is $a=w/2$. To describe the modification of the front temperature gradient near the ice–water boundary, we approximate the front cap as a circular liquid inclusion of radius $a$ embedded in a solid half-space [Fig.~\ref{figS2}(b)]. The external ice–water boundary is represented by the plane $x=0$, with $x>0$ directed into the solid, and the front-interface separation is denoted by $d_\mathrm{f}$. The circular inclusion is therefore centered at $(x,y)=(d_\mathrm{f}+a,0)$. 

Assuming quasi-steady, conduction-dominated heat transfer, the temperature fields in the solid and liquid satisfy
\begin{equation}
    \nabla^2 T_\mathrm{s}=0,
    \quad
    \nabla^2 T_\mathrm{l}=0.
    \tag{S1}
    \label{S3}
\end{equation}

The ice-water boundary is taken to be isothermal, $T_\mathrm{s}=T_\mathrm{i}$ at $x=0$, and far from the inclusion, the distorting effect of the inclusion becomes weak and the imposed temperature gradient is recovered, $T_\mathrm{s}\sim T_\mathrm{i}-G_\mathrm{s} x$. Temperature and normal heat flux are taken to be continuous across the circular solid--liquid interface $\Gamma$,
\begin{equation}
    T_\mathrm{s} = T_\mathrm{l},
    \qquad
    k_\mathrm{s}\frac{\partial T_\mathrm{s}}{\partial n}
    =
    k_\mathrm{l}\frac{\partial T_\mathrm{l}}{\partial n},
    \qquad \bm{x}\in\Gamma.
    \tag{S2}
    \label{S4}
\end{equation}

The geometry is conveniently represented in bipolar coordinates $(\xi,\eta)$:
\begin{equation}
    x=c\frac{\sinh\xi}{\cosh\xi-\cos\eta},\quad
    y=c\frac{\sin\eta}{\cosh\xi-\cos\eta}.
    \tag{S3}
    \label{S5}
\end{equation}

The planar boundary and circular inclusion correspond to $\xi=0$ and $\xi=\xi_{0}$, respectively. Matching the radius and center position of the inclusion gives $c=a\sinh\xi_0$ and $\cosh\xi_0=1+d_\mathrm{f}/a$. The solid occupies $0<\xi<\xi_0$, while the interior of the liquid inclusion corresponds to $\xi>\xi_0$.

Introducing the normalized temperature potential $u={(T_i-T)}/{G_s}$, the undisturbed temperature field is simply $u_{\mathrm{bg}}=x$. We therefore write the solid-side solution as $u_{\mathrm{s}}=x+\widetilde{u}_s$, where $\widetilde{u}_s$ shows the distortion generated by the circular conductivity discontinuity. The disturbance therefore satisfies
\begin{equation}
    \widetilde{u}_{\mathrm{s},\xi\xi}
    +
    \widetilde{u}_{\mathrm{s},\eta\eta}
    =0,
    \quad
    \widetilde{u}_\mathrm{s}(0,\eta)=0 \quad(\eta\neq0).
    \tag{S4}
    \label{S6}
\end{equation}

Symmetry about $y=0$ and periodicity in $\eta$ allow the solutions to be expanded in cosine modes. Imposing the planar boundary condition and boundedness inside the liquid inclusion gives 

\begin{align}
{u}_\mathrm{s}
    =
    x
    +
    \sum_{n=1}^{\infty}
    A_n\sinh(n\xi)\cos(n\eta),
       \tag{S5-1}
    \label{S7-1}
    \\
    u_\mathrm{l}
    =
    c
    +
    \sum_{n=1}^{\infty}
    B_n
    e^{-n(\xi-\xi_0)}
    \cos(n\eta).
       \tag{S5-2}
    \label{S7-2}
\end{align}

We define the conductivity ratio $K = k_\mathrm{l}/k_\mathrm{s}$. Matching temperature and normal heat flux at $\xi=\xi_0$ yields

\begin{equation}
    A_n
    =
    \frac{
        2ce^{-n\xi_0}(1-K)
    }{
        \cosh(n\xi_0)+K\sinh(n\xi_0)
    },
    \quad
    B_n
    =
    \frac{
        2c
    }{
        \cosh(n\xi_0)+K\sinh(n\xi_0)
    }.
    \tag{S6}
    \label{S8}
\end{equation}

The front tip corresponds to $(\xi,\eta)=(\xi_0,\pi)$ and along the symmetry axis $\eta=\pi$,
\begin{equation}
    x
    =
    c\tanh\left(\frac{\xi}{2}\right),
    \qquad
    \frac{\partial x}{\partial\xi}
    =
    \frac{c}{\cosh\xi+1}.
    \tag{S7}
    \label{S9}
\end{equation}

The normalized liquid-side temperature gradient at the front is therefore
\begin{equation}
    \chi_\mathrm{f}
    \equiv
    \frac{G_\mathrm{f}}{G_\mathrm{s}}
    =
    \left.
    \frac{\partial u_\mathrm{l}}{\partial x}
    \right|_{\xi=\xi_0,\eta=\pi}
    =
    2(\cosh\xi_0+1)
    \sum_{n=1}^{\infty}
    \frac{
        (-1)^{n+1}n
    }{
        \cosh(n\xi_0)+K\sinh(n\xi_0)
    }.
    \tag{S8}
    \label{S10}
\end{equation}

For numerical evaluation, we use the equivalent rapidly convergent form
\begin{equation}
    \chi_\mathrm{f}=
\frac{2}{1+K}
\sum_{m=0}^{\infty}
\left(-\frac{1-K}{1+K}\right)^m
\frac{\cosh^2(\xi_0/2)}
{\cosh^2[(2m+1)\xi_0/2]} .
   \tag{S9}
    \label{S11}
\end{equation}

The series is evaluated adaptively until an upper bound on the truncation error falls below $10^{-12}\max(1,|S_m|)$, where $S_m$ is the accumulated partial sum.

\subsection*{Analytical rear-end conduction solution}
To describe the rear-end thermal field after connection to the external liquid, we approximate the local geometry by a circular liquid region of radius $a=w/2$ intersecting a planar ice–water boundary at $x=0$ [Fig.~\ref{figS2}(c)]. The circle is centered at $C=(h,0)$ and satisfies $(x-h)^2+y^2=a^2$, where $h=d_\mathrm{r}-a$ and $d_\mathrm{r}$ is the distance from the rear tip to the planar boundary. Introducing an intersection angle $\theta$, the geometry is specified by
\begin{equation}
d_\mathrm{r}^*\equiv\frac{d_r}{a}=1+\cos\theta,\quad
b=a\sin\theta,\quad 0<\theta<\pi,
   \tag{S10}
   \label{S12}
\end{equation}
where $2b$ is the opening width. For $0<d_\mathrm{r}^*<2$, the intersected-circle geometry serves as a local approximation to the connected liquid finger.

The domain is mapped using rotated bipolar coordinates, in which the two intersection points $P_{\pm}$ are the foci of this system,
\begin{equation}
x=
a\sin\theta\,
\frac{\sin\eta}{\cosh\xi-\cos\eta},
\quad
y=
a\sin\theta\,
\frac{\sinh\xi}{\cosh\xi-\cos\eta},
   \tag{S11}
   \label{S13}
\end{equation}
where the solid and liquid occupy $0<\eta<\theta$ and $\theta<\eta<2\pi$, respectively. The circular solid–liquid interface $\Gamma_c$ corresponds to $\eta=\theta$, while the remaining planar interface $\Gamma_p$ is represented by $\eta=0$ on the solid side and $\eta=2\pi$ on the liquid side. The liquid opening corresponds to $\eta=\pi$ and does not constitute a solid–liquid boundary.

Assuming quasi-steady conduction, the temperatures in the solid and liquid domains satisfy Eq.(\ref{S3}). Across the ice-water interface, temperature and conductive normal heat flux are taken to be continuous, 
\begin{equation}
T_\mathrm{s} = T_\mathrm{l},
\quad
k_\mathrm{s} \nabla T_\mathrm{s} \cdot \mathbf{n}
=
k_\mathrm{l} \nabla T_\mathrm{l} \cdot \mathbf{n},
\quad
\mathbf{x} \in \Gamma_p \cup \Gamma_c .
   \tag{S12}
   \label{S14}
\end{equation}

Similarly, we define the conductivity ratio $K={k_\mathrm{l}}/{k_\mathrm{s}}$ and introduce the temperature potentials $u_j=(T_\mathrm{i}-T_j)/{G_\mathrm{s}}$, where the subscripts s and l denote the solid and liquid, respectively.

To decompose the temperature potentials into a background conduction field and a geometry-induced perturbation, we choose
\begin{equation}
u_\mathrm{s}=x+v_\mathrm{s},
\quad
u_\mathrm{l}=\frac{x}{K}+v_\mathrm{l}.
   \tag{S13}
   \label{S15}
\end{equation}

Here, $v_j$ has a Fourier cosine integral representation
\begin{equation}
v_j(\xi,\eta)
=\int_0^\infty
\widehat{v}_j(q,\eta)
\cos(q\xi)\,dq,\quad j=\mathrm{s},\mathrm{l}.
   \tag{S14}
   \label{S16}
\end{equation}

The modal functions are then given by
\begin{equation}
\widehat{v}_s(q,\eta)
=
\alpha_s(q)\cosh(q\eta)
+
\beta_s(q)\sinh(q\eta),
\quad
0<\eta<\theta,
   \tag{S15}
   \label{S17}
\end{equation}
and
\begin{equation}
\widehat{v}_l(q,\eta)
=
\alpha_l(q)\cosh(q\eta)
+
\beta_l(q)\sinh(q\eta),
\quad
\theta<\eta<2\pi.
   \tag{S16}
   \label{S18}
\end{equation}
The four modal amplitudes are determined by imposing temperature and heat-flux continuity at the planar interface $\eta=0,2\pi$ and the circular interface $\eta=\theta$, yielding the four matching conditions
\begin{align}
\hat v_s(q,0) = \hat v_l(q,2\pi),
\tag{S17-1}
\label{S19-1}
\\
\partial_\eta \hat v_s(q,0)= K\partial_\eta \hat v_l(q,2\pi),
\tag{S17-2}
\label{S19-2}
\\
\hat v_s(q,\theta)-\hat v_l(q,\theta)= J(q),
\tag{S17-3}
\label{S19-3}
\\
\partial_\eta \hat v_s(q,\theta)
= K\partial_\eta \hat v_l(q,\theta).
\tag{S17-4}
\label{S19-4}
\end{align}
The inhomogeneous term $J(q)$ satisfies
\begin{equation}
J(q)=2a\sin\theta
\left(\frac{1}{K}-1\right)
\frac{\sinh[q(\pi-\theta)]}{\sinh(\pi q)}.
\tag{S18}
\label{S20}
\end{equation}

Based on this, evaluating the liquid-side gradient at the rear tip, $(\xi,\eta)=(0,\theta)$, yields
\begin{equation}
\chi_\mathrm{r}(K,\theta)
=
\frac{1}{K}+
2(1-\cos\theta)\left(1-\frac{1}{K}\right)
\int_0^\infty\frac{q\sinh[q(\pi-\theta)]H(q)}{\sinh(\pi q)\Delta(q)}dq,
 \tag{S19}
   \label{S21}
\end{equation}
with
\begin{align}
H(q)=\cosh(q\theta)\sinh[q(2\pi-\theta)]+\frac{1}{K}\sinh(q\theta)\cosh[q(2\pi-\theta)],
    \tag{S20-1}
    \label{S22-1}
\\
\Delta(q)=4\sinh^2(\pi q)+\frac{(1-K)^2}{K}\sinh(q\theta)\sinh[q(2\pi-\theta)].
    \tag{S20-2}
    \label{S22-2}
\end{align}

All quantities in Eq.~\eqref{S21} are explicit functions of the conductivity ratio $K$ and the geometric intersection angle $\theta$, where $\theta=\cos^{-1}(d_r^*-1)$. The experimentally relevant rear-gradient correction is therefore $\chi_r(d_r^*;K)=\chi_r \left[K,\cos^{-1}(d_r^*-1)\right]$.

\begin{figure*}[!htb]
\centering
\includegraphics[width=1\linewidth]{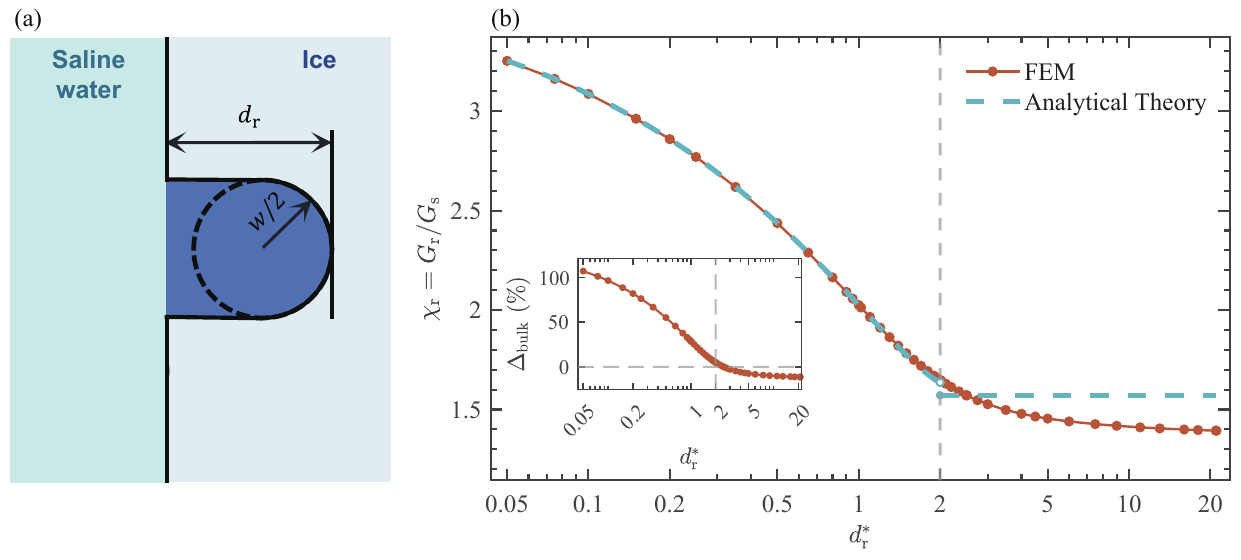}
\caption{Geometric model and thermal-gradient correction at the rear end of a connected brine pocket. (a) Idealized geometry of the connected liquid finger used in the finite-element calculations. The finger has a width $w$ and terminates in a semicircular rear cap of radius $w/2$, connecting the embedded liquid to the external saline water through an opening in the planar ice–water interface. The distance from the interface to the rear tip is denoted by $d_\mathrm{r}$. (b) Normalized liquid-side temperature gradient at the rear tip, $\chi_\mathrm{r}=G_\mathrm{r}/G_\mathrm{s}$, as a function of the normalized rear-tip distance $d_\mathrm{r}^*=2d_\mathrm{r}/w$. Orange symbols represent FEM results, while the cyan dashed curve represents the analytical reference, comprising the intersected-circle solution for $d_\mathrm{r}^*<2$ and the bulk approximation, $\chi_{\mathrm{bulk}}=2/(1+K)$, for $d_\mathrm{r}^*\geq 2$, where $K=k_\mathrm{l}/k_\mathrm{s}$ is the liquid-to-ice thermal conductivity ratio. The vertical dashed line marks the transition at $d_\mathrm{r}^*=2$. The inset shows the relative deviation of the FEM results from the bulk approximation, $\Delta_{\mathrm{bulk}}=100(\chi_\mathrm{r}/\chi_{\mathrm{bulk}}-1)$.
}
\label{figS3}
\end{figure*}

Eq.~\eqref{S21} describes the rear-gradient correction when the planar interface intersects the circular cap, $0<d_\mathrm{r}^*<2$. When the front first connects to the external liquid, the rear tip can still lie at $d_\mathrm{r}^*>2$, with an extended liquid finger linking the rear cap to the opening. To determine the thermal correction during this earlier connected stage and assess the bulk approximation used in the main text, we solve the steady two-material conduction problem numerically. The finger consists of a straight liquid section of width $w=2a$ and length $L=d_\mathrm{r}-a$, terminated by a semicircular cap of radius $a$ [Fig.~\ref{figS3}(a)]. We extract the liquid-side tip gradient $G_{\mathrm{r}}$ and normalize it as $\chi_\mathrm{r}=G_{\mathrm{r}}/G_{\mathrm{s}}$.

We solve $\nabla\!\cdot(k\nabla T)=0$ with the boundary temperatures and layer thicknesses specified in the Methods. Temperature and normal heat flux are continuous across the phase boundary, and the transverse boundaries have zero normal heat flux. The corresponding planar two-layer solution defines $G_{\mathrm{s}}$. For $0<d_{\mathrm r}^*<1$, the numerical geometry reduces to the intersected circle.

Fig.~\ref{figS3}(b) compares the FEM results (orange circles joined by a solid line) with the piecewise theoretical prediction used in the main text (cyan dashed line), comprising Eq.~\eqref{S21} for $0<d_{\mathrm r}^*<2$ and the bulk expression $\chi_{\mathrm{bulk}}=2/(1+K)$ for $d_{\mathrm r}^*\geq2$. The vertical dashed lines in the main panel and inset mark $d_{\mathrm r}^*=2$, corresponding to $d_{\mathrm r}=w$ and the transition between the two theoretical branches. Over the sampled positions with $d_{\mathrm r}^*<2$, the numerical results agree with Eq.~\eqref{S21} to within $1\%$, supporting the circular-cap approximation near the interface.

The inset quantifies the departure from the bulk approximation through
\[
\Delta_{\mathrm{bulk}}
=
\left(
\frac{\chi_{\mathrm r}^{\mathrm{FEM}}}{\chi_{\mathrm{bulk}}}-1
\right)\times100\%,
\]
with the horizontal dashed line indicating zero deviation. For long fingers, the extended slab-like liquid section modifies the temperature field at the rear cap, yielding a gradient slightly below the bulk value at the longest simulated geometry. This reduction is appreciable primarily for pockets with large initial lengths and diminishes as the finger shortens. It is small relative to the pronounced positive enhancement as $d_{\mathrm r}^*<2$ and is neglected in the main-text piecewise approximation. Together, the main panel and inset show that strong rear-gradient enhancement develops only when the rear approaches within a distance comparable to the pocket width. This behavior supports the thermal correction used in Fig.~2(b) of the main text and is consistent with the finding that thermal-field modification alone does not reproduce the earlier onset of the measured rear acceleration.

\subsection*{MOVIE CAPTIONS}
\textbf{Movie.S1} Color-coded segmentation and tracking of brine pockets during ice aging under the experimental conditions described in Fig.~1. Each pocket is highlighted by a semitransparent segmentation mask and its trajectory, shown in the same color. Colors remain consistent throughout the sequence. Elapsed time is indicated in the lower-left corner, with an interval of 10 s between successive frames.

\textbf{Movie.S2} Original image sequence corresponding to Movie.S1. The same image sequence is shown without segmentation masks or trajectories, allowing direct comparison with Movie S1. Elapsed time is indicated in the lower-left corner, with an interval of 10 s between successive frames.
\bigskip
\noindent

%


\end{document}